
\font\titlefont = cmr10 scaled \magstep2
\magnification=\magstep1
\vsize=22truecm
\voffset=1.75truecm
\hsize=15truecm
\hoffset=0.95truecm
\baselineskip=20pt

\settabs 18 \columns

\def\b{\bigskip}
\def\bb{\bigskip\bigskip}

\def\ce{\centerline}

\def\no{\noindent}

\rightline{AMES-HET-95-14}
\rightline{December 1994}
\bb

\ce{\titlefont{Non-Universal Correction To $Z \rightarrow b
    \overline{b}$  }}
\ce{\titlefont{And Flavor Changing Neutral Current Couplings}}

\b

\ce{ X. Zhang ~and ~B.-L. Young}
\b
\ce{Department of Physics and Astronomy }
\ce{ Iowa State University,}
\ce{ Ames, Iowa
 50011}

\b
\bb
\ce{\bf ABSTRACT}
  A non-universal interaction, which involes
only the heavy quarks
 $(t_L, ~b_L)$ and
$t_R$, modifies the neutral current couplings and
induces flavor changing neutral currents (FCNC).
The size of the FCNC effect depends crucially on the dynamics of the fermion
mass generation.
In this paper, we study the
  effect of the non-universal interaction
 on $Z b
{\overline{b}}$, $Z b {\overline{s}}$,
 $Z d {\overline{s}}$ and
$Z d {\overline{b}}$, by using
an effective lagrangian technique and
assuming the quark
mass matrices in the form of a
generalized Fritzsch ansatz. We point out that
if fitting $R_b = \Gamma (Z \rightarrow
b {\overline{b}}) / \Gamma( Z \rightarrow
{\rm Hadrons} )$ to the LEP data within 1 $\sigma$,
the induced FCNC couplings are very
close to the allowed bounds of several
rare decays.

\bb
\filbreak

Recently
the CDF collaboration[1] at FNAL presented evidence for a top quark with a mass
$m_t \sim 175$ GeV. Since
$m_t$ is of the order of Fermi scale, the top quark couplies strongly to
the electroweak symmetry breaking sector and
 will play a key role in
probing new physics beyond the standard model. This kind of new physics
 ( {\it i.e.} non-universal
interaction since it acts on only the top quark)
 can become manifest in top quark production
processes at the hadron and next
generation linear colliders.
It can also affect the partial width of
$Z \rightarrow b {\overline{b}}$
measured
at LEP because the $SU(2)_L$ group places $(t_L, ~ b_L)$ into a common
doublet.
The experimental observed value for
  the ratio $R_b = \Gamma(Z \rightarrow b
{\overline{b}} ) / \Gamma (Z \rightarrow {\rm Hadrons} )$ is slightly higher
than the standard model expectation. This
may be an indication of the non-universal interaction,
if it is more than a statistical fluctuation.

It
is known that a
 non-universal interaction will induce
flavor changing neutral currents (FCNC) among the light fermions[2-5].
 However,
the size of the FCNC effect depends crucially on
the quark mass mixing matrices. So one can not
predict quantatively the induced FCNC effect without
specifying the mass matrices.
 At present it seems far too early to attempt
an actual solution to the issue of mass
generation.
However,
there has been a great amount of activities in looking for
the relation between fermion masses and their mixing matrix elements,
as commonly
refered to as
texture studies.
One expects that
 a ``successful" ansatz
can
 provide clues to the dynamics
 of the fermion mass generation.

In recent years, most studies
 on the implication of fermion mass ansatz were focused on
 grand unification theories with and without supersymmetry.
 In this paper we
take a phenomenological, model independent approach
to
 new physics beyond the standard model, {\it i.e.}, the
 effective lagrangian technique, and
consider the implication of the fermion mass ansatz on the
induced FCNC effect.
Specifically, we will use one variation of the
 Fritzsch[6] ansatz to
 study the correlated effects of new physics on
$Z b{\overline{b}}$ and
$Z b {\overline{s}}$, {\it etc}.
 We will point out that when fitting
 $R_b$
  to the LEP data within $1 \sigma$,
the induced FCNC couplings are very close to the allowed
bounds of several rare decays. Our results
show that the new physics associated
with top quark may be revealed by
the presence of FCNC processes.

We first
 discuss $Z \rightarrow b {\overline{b}}$.
Following the general approach, we
assume that anomalous, non-universal interaction is
$SU(2)_L \times U(1)_Y$ invariant. Hence the $b$ quark will participate in
any $t$ quark interactions when the left-handed doublet is involved. This
 can result in a modification
of
the $Z b {\overline{b}}$
vertex. We can parametrize the modification by
introducing a parameter $\kappa_j$,
which shifts the standard model tree level coupling,
$g_{j}$, to effective coupling $g^{eff}_{j}$;

$${
g_j^{eff} = g_j (1 + \kappa_j) ~~,
} \eqno(1.a)$$

\no where $j= L (R)$ denotes the left-(right) hand, and $g_j$
are the standard model coupling strengths of the neutral current,

$${
g_L = -{1 \over 2} + {1\over 3} \sin^2\theta_W ~~; ~~~~~
 g_R = {1\over 3} \sin^2 \theta_W ~~. } \eqno(1.b)$$

\no The contributions of the new physics
to the $Z \rightarrow b {\overline{b}}$
width
 are proportional to $g_L^2$ and
$g_R^2$. Since
$g^2_L >> g^2_R$,
 we will neglect the modification to
the right-handed interaction in this article.
Defining $\delta \Gamma$ to be the purely non-universal
correction of the new physics beyond the standard
 model to the $Z \rightarrow b {\overline{b}}$ width,
$\Gamma_{b {\overline{b}}}$, we have
$${
 {\delta \Gamma \over \Gamma_{b {\overline{b}}} } \simeq 2~
        { g_L^2 \kappa_L  \over {g_L^2 + g_R^2 } } \simeq 2 ~ \kappa_L
 ~~~. } \eqno(2)$$

\no Then the $R_b$ becomes

$${
R_b \sim R_b^{SM} \left( 1 + {\delta \Gamma \over \Gamma_{b {\overline{b}}} }
\right) \sim R_b^{SM} ( 1 + 2 \kappa_L ) ~ ,  } \eqno(3)$$

\no where the standard model value $R_b^{SM} =0.2157$ for
$m_t = 175 $GeV and
$m_H = 300$ GeV. The experimental value
of $R_b$ measured at LEP is
$R_b = 0.2192 \pm 0.0018$[7], which is roughly
within 2$\sigma$ of the standard model expectation.
A positive $\kappa_L$ would improve the situation.

In general, $\kappa_{L}$ can be viewed as functions of $q^2$[8],
where $q$ is the 4-momentum of the Z-boson, and at LEP, $q^2 = m_Z^2$.
Expanding $\kappa_{L}$ in terms of $q^2$, we
have
$${
\kappa_{L} = \kappa_{L}^0 +  q^2-{\rm dependent ~terms }
{}~~. }\eqno(4)$$

\no Gauge invariant operators describing $\kappa_{L}$
have been constructed explicitly in effective
lagrangian with a
non-linear[2] realization of $SU(2)_L \times U(1)_Y$.
In this paper we use an effective lagrangian with a
 linear
realization[9] of
$SU(2)_L \times U(1)_Y$ for the discussion.
The new physics effects are parametrized by a set of higher dimension
operators $ {\cal O}^i $, which are
required to be invariant under the standard model
gauge symmetry and contain only the standard model fields.
The new physics effects on the light fermions are
assumed to be negligible, so
the higher dimension operators involve only $( t_L, ~ b_L ), ~~t_R$,
the gauge and scalar bosons. For dimension 6,
there are two operators which
generate directly
 \footnote{[F.1]}{
We are not considering the
operators which can affect $Z b {\overline{b}}$
indirectly by loop effects.}
 a $\kappa_{L}^0$ in eq.(4)[10],

$${
{\cal O}^1 = i~ [\phi^\dagger D_\mu \phi - {( D_\mu \phi )}^\dagger \phi ]
               ~{ \overline{\Psi}}_L \gamma^\mu \Psi_L ~~; }\eqno(5.a)$$

$${
{\cal O}^2 = i~ [ \phi^\dagger
  { {\vec{\tau}} }
 D_\mu \phi - {( D_\mu \phi )}^\dagger
{ {\vec{\tau}}}
 \phi ]   ~
               { \overline{\Psi}}_L \gamma^\mu { {\vec{\tau}} }
                      \Psi_L
{}~~, } \eqno(5.b)$$

\no where $\phi$ is the doublet Higgs field
of the standard model and $\Psi_L^T = (t, b)_L$.
 Let us introduce the effective lagrangian, ${\cal L}^{eff}$,
containing higher dimension operators given in eqs.(5):

$${
{\cal L}^{eff} = {\cal L}^{SM} + {1 \over \Lambda^2} \left(
         c_1 {\cal O}^1 + c_2 {\cal O}^2  \right)
{}~~, } \eqno(6)$$

\no where $c_i , ~ i= 1, 2$, are real parameters,
which determine the strength of
the contributions of the operators,
${\cal L}^{SM}$ is the standard model
lagrangian, $\Lambda$ is the cutoff of the effective theory.

After the electroweak symmetry breaking, the anomalous couplings
for $Z b {\overline{b}}$ and
$Z b {\overline{s}}$, {\it etc.},
from ${\cal L}^{eff}$ are contained in

$${
{g \over { \cos\theta_W} } {{ \pmatrix{{\overline d}
  \cr {\overline s} \cr
{\overline
b} \cr}}}^T_L
   U_L^{ \dagger (d)} \pmatrix{ 0 & & \cr
                                     & 0 & \cr
                                    & &  \delta_L \cr } U_L^{(d)}\gamma_\mu
                                    \pmatrix{d \cr s \cr b \cr}_L ~Z^\mu
{}~~, } \eqno(7) $$

 \no where

$${
\delta_L = {v^2 \over \Lambda^2 } ~~( c_1 +  c_2 ) ~~; }\eqno(8)$$

\no and $v \simeq 250$ GeV,
$U_{L}^{(d)}$ is unitary rotation matrix on the left-handed
 down quarks.
The Cabibbo-Kabayashi-Maskawa (CKM) mixing matrix for the charged weak current
is

$${
V = {( U_L^{(u)} )}^\dagger U_L^{(d)}
{}~~~,  } \eqno(9) $$
\no where $U_L^{(u)}$ is the unitary rotation matrix for the left-handed up
quarks.
 Note that in the standard model,
which corresponds to ${\cal L}^{eff}$ in the limit
$\Lambda \rightarrow \infty$,
  the individual $U_{L, R}^{(u)}$
and $U_{L, R}^{(d)}$ are not measureable, but only $V$ in eq.(9) is.
Furthermore,
the universality of the weak interaction in the standard model also guarantees
the vanishes of the FCNC at tree level.

The relative size of the
$Z b {\overline{b}}$ to the FCNC couplings,
$Z b {\overline{s}}$, {\it etc.}, in eq.(7)
depends on the rotation matrix $U_{L}^{(d)}$.
The elements of
 $U_{L}^{(d)}$ can be evaluated once
the corresponding mass matrix
is given. In
the literature a widely
used ansatz is the one suggested by Fritzsch[6] and its variations.
 The latter is given by
\b
$${
M^{(q)} = \pmatrix{ 0 & x_q e^{i \alpha_q} & 0 \cr
                    x_q e^{-i \alpha_q} & \omega_q & y_q e^{ i \beta_q } \cr
                   0 & y_q e^{- i \beta_q} & z_q \cr }
{}~~, ~} \eqno(10)$$
\b

\no where $x_q, ~ y_q,~ \omega_q$ and
$z_q$ are real parameters and
$q=u ~(d)$ denotes the up (down) type quarks.
The original Fritzsch ansatz is given by putting $\omega_q = 0$, which
predicts a too small top quark mass $m_t \leq 90$ GeV[11].
 Here we consider one
variation[12] which
 can have an acceptable top quark mass $m_t \leq 190$ GeV, and
 fits the
current experimental data on the CKM matrix. In
the variation[12], the rotation matrix,
$U^{(q)}$ (= $U^{(q)}_L = U^{(q)}_R$ ) is given by
\b
$${
\pmatrix{ 1 & -{\left( {m_1 \over m_2 } \right)}^{1/2} & {\left( {
              m_1 m_2 (m_2 + w_q) \over m_3^3 } \right) }^{1/2} \cr
         {\left( {m_1 \over m_2} \right) }^{1/2} e^{-i \alpha_q}
                  & e^{-i \alpha_q} & { \left( {m_2 + w_q \over m_3 }
                     \right)}^{1/2} e^{-i \alpha_q} \cr
        -{ \left( {m_1( m_2 + w_q ) \over {m_2 m_3 }} \right) }^{1/2}
           e^{-i (\alpha_q + \beta_q )} &
        - { \left( {m_2 + w_q \over m_3 } \right) }^{1/2} e^{-i( \alpha_q
                       + \beta_q )} & e^{-i ( \alpha_q + \beta_q )}
                  \cr }
 }\eqno(11)$$

\b
\no where $m_1,~ m_2, ~{\rm and}~ m_3$ correspond to $m_u, ~ m_c ~{\rm and}~
m_t$ for
$q=u$, and $m_d, ~ m_s~$ and
 $m_b$ for
$q=d$, and
$w_u = m_c, ~~w_d = 0$, $\alpha_q ~{\rm and}~ \beta_q$ are responsible for
the CP violation phase in the CKM matrix.

 In table I, we give the theoretical values of FCNC couplings
and the corresponding experimental upper limits.
 One can see that if fitting $R_b$ to LEP data
within 1$\sigma$,
 the induced FCNC couplings
 are close to the allowed bounds of several
rare decays\footnote{[F.2]}
{We realize
 that there are uncertainties in the numerical values of the rotation
matrix elements caused by the uncertainties in the
values of fermion masses, CKM mixing
angles and the analytical approximation used in ref.[12].}.
 For example, assuming a
 positive $\kappa_L$ and
fitting $R_b$ to the experimental data within $1 \sigma$, we have
$| {\tilde \kappa}_L^{ds} | \geq ~(1.2 \sim 2.6) \times 10^{-5}$, which
lies in the
experimental limit of $K_L \rightarrow {\overline \mu} \mu$.

In our calculations we have not considered
  the $q^2$ dependent terms in
eq.(4), which are generally proportional to $m^2/ \Lambda^2$ where
$m$ is a typical mass of a process under consideration. The operators in
eqs.(5)
 give rise to terms proportional to $v^2 / \Lambda^2$. Therefore the
momentum dependent terms are generally suppressed at low energies.
           We should point out that
 if a different ansatz from that in (11)
 is taken,
the relative size of
anomalous $Z b {\overline{b}}$ to
$Z b {\overline{s}}$, {\it etc.}, may be changed.
Thus the future data on $Z b {\overline{b}}$ and $Z b {\overline{s}}$,
{\it etc.}, will provide an
experimental test on various fermion mass ans\"atz.

\b
\bb
XZ is grateful to G. Valencia for discussions and to Zhi-Zhong Xing
for useful correspondence on the Fritzsch ansatz.
This work is supported in part by the Office of High Energy
 and Nuclear Physics
of the U.S. Department of Energy (Grant No. DE-FG02-94ER40817).

\bb
\vfill\eject
\b
\ce {\bf {Table Caption}}

\no {\bf {Table I:}}
Theoretical prediction on FCNC couplings, and corresponding
experimental upper limits taken
from Ref.[13]. ${\tilde{\kappa}}_{L} =
 U^{\dagger (d)}_{L}~
{\it diag}[ 0, 0, \delta_{L} ]~ U_{L}^{(d)} $. The
elements of $U_{L}^{(d)}$ are calculated by taking the central values of
the down quark masses evaluated at $\mu =$ 1 GeV,
$m_s/m_b = 0.033, ~~ m_d/m_s = 0.051$.
For $Z \rightarrow b {\overline{b}} ,
{}~ {\tilde \kappa}_L^{bb} = \delta_L$,  and using
definition of $\kappa_L$ in eq.(1.a) we have $\delta_L = g_L \kappa_L,
{}~~{\rm so}
 ~  { R_b-R_b^{SM} \over
R_b^{SM}} = { 2 \delta_{L} \over
g_L} $.
 \b

\bb
\b
\vbox{\tabskip=0pt \offinterlineskip
\def\tablerule{\noalign{\hrule}}
\halign to300pt {\strut#& \vrule#\tabskip=1em plus2em&
  \hfil#& \vrule#& \hfil#\hfil& \vrule#&
    \hfil#& \vrule# \tabskip=0pt \cr \tablerule

&& $ |{\tilde{\kappa}}_{L}^{ij} |$ && Predictions   && Limits and
                                                   Processes
                                                               &\cr \tablerule
&&  $ | {\tilde{\kappa}}_{L}^{ d s} | $ && $ |7.5 \times 10^{-3}~
  \times  \delta_{L} |$   &&  $3 \times 10^{-4}$ ($K^0 - {\overline{K^0}}$
mixing)   &\cr \tablerule
&& $ |{\tilde{\kappa}}^{ds}_{L}| $ &&
$ | 7.5 \times 10^{-3}~ \times \delta_{L} | $ &&
$2 \times 10^{-5}$  ( $ K_L \rightarrow {\overline{\mu}} \mu$ )
 &\cr \tablerule
&& $ |{\tilde{\kappa}}_{L}^{db}| $  && $ |
0.041  \times
 \delta_{L} | $ && $4 \times 10^{-4}$  ( $B_d - {\overline{B}}_d
$ mixing) &\cr \tablerule
&& $|{\tilde{\kappa}}_{L}^{bs}|$ &&  $ | 0.182 \times \delta_{L} | $
 &&   $2 \times 10^{-3}$   ( $B \not\rightarrow
l^+ l^- X$ ) &\cr \tablerule
 \noalign{\smallskip}
& \multispan{7} Table I. \hfil\cr}}

\vfill\eject
\ce {\bf References}

\item{[1]}F. Abe, et al., Phys. Rev. {\bf D50}, 2966 (1994).

\item{[2]}R.D. Peccei and X. Zhang, Nucl. Phys. {\bf B337}, 269 (1990).

\item{[3]}C.T. Hill, FERMILAB-PUB-94/395-T, November (1994).

\item{[4]}B. Holdom, Preprint of University of Toronto, UTPT-94-20,
            hep-ph/9407311.

\item{[5]}H. Georgi, L. Kaplan, D. Morin, A. Schenk, Harvard
 University Preprint, hep-ph/9410307.

\item{[6]}H. Fritzsch, Phys. Lett. {\bf 73B}, 317 (1978); Nucl.
 Phys. {\bf B155}, 189 (1979).

\item{[7]}R. Batley, talk presented at DPF `94, Albuquerque, New Mexico,
Aug. 2-6, 1994.

\item{[8]}C.T. Hill and X. Zhang, FERMILAB-PUB-94/231-T, July (1994),
to appear in Phys. Rev. {\bf D}.

\item{[9]}For example, see, W. Buchm\"uller, D. Wyler, Nucl. Phys. {\bf268},
621 (1986).

\item{[10]}The operator ${\cal O}^1 ~{\rm and}~
{\cal O}^2$ modify also $Z t_L {\overline{t}}_L$ and
$W t_L {\overline{b}}_L$ vertices.
There exists a operator similar to ${\cal O}^1$ with $\Psi_L$ being
replaced by
$t_R$. This third operator gives rise to
a correction to the $Z t_R {\overline{t}}_R$ vertex. A full
 discussion
on the implication of the FCNC effects,
such as $Z t {\overline{c}}$,
 associated with these anomalous
gauge couplings will be presented in a future publication.

\item{[11]} C.H. Albright, B.A. Lindholm and C. Jarlskog, Phys. Rev. {\bf D38},
872 (1988); Y. Nir, SLAC Report No. SLAC-PUB-5676 \& WIS-91/73/Oct-PH, 1991.

\item{[12]}Dongsheng Du and Zhi-Zhong
 Xing, Phys. Rev. {\bf D48}, 2349 (1993), and
references theirin.

\item{[13]}C.P. Burgess et al., Phys.
Rev. {\bf D49}, 6115 (1994);
E.W.J. Glover and
J.J. van der Bij, in {\it Z-physics at LEP I},
 CERN yellow report CERN 89-08, Vol. 2, P.42,  ed. G. Altarelli,
 R. Kleiss and C. Verzegnassi.

\bb
\b

\bye

\bye